# Femtosecond Engineering of magnetic Domain Walls via Nonequilibrium Spin Textures

Yuzhu Fan[1], Gaolong Cao[1], Junlin Wang[2], Sheng Jiang[3], Jing Wu[2,4], Johan Åkerman[5,6,7], Jonas Weissenrieder[1]

[1]*School of Engineering Sciences, KTH Royal Institute of Technology, Applied Physics, AlbaNova, SE-106 91 Stockholm, Sweden*

[2]*School of Integrated Circuits, Guangdong University of Technology, 510006 Guangzhou,,China*

[3]*School of Microelectronics, South China University of Technology, 510641 Guangzhou, China*

[4]*School of Physics, Electronics and Technology, University of York, York, UK*

[5]*Department of Physics, University of Gothenburg, Gothenburg, Sweden.*

[6]*Center for Science and Innovation in Spintronics, Tohoku University, 2-1-1 Katahira, Aoba-ku, Sendai, Japan.*

[7]*Research Institute of Electrical Communication, Tohoku University, 2-1-1 Katahira, Aoba-ku, Sendai, Japan.*

**Abstract**

Ultrafast optical control of magnetic textures offers new opportunities for energy-efficient, high-speed spintronic devices. While uniform magnetization reversal via all-optical switching is well established, the formation dynamics of non-uniform domain walls (DWs) under ultrafast excitation remain poorly understood. Here, we use Lorentz ultrafast electron microscopy combined with transient optical grating excitation to directly image the real-time formation of DWs in a ferrimagnetic GdFeCo film. We observe a rapid evolution from disordered spin contrast to ordered DW arrays within 10 ps, including a transient, strongly asymmetric DW state. In a narrow fluence window, short-lived DWs form and spontaneously vanish within picoseconds. Multiscale simulations combining atomistic spin dynamics and micromagnetics reveal a nonlinear nucleation pathway involving a hybrid transition state where localized, unstable spin textures coalesce into metastable DWs. This nonequilibrium mechanism explains the observed asymmetry and spatial ordering, and establishes a framework for controlling spin textures in magnetic materials on femtosecond timescales.

**Keywords:** all optical switch, magnetic domain wall engineering, ultrafast Lorentz transmission electron microscopy, nonequilibrium magnetic dynamics

**Main**

The ability to control magnetism on ultrafast timescales is opening new frontiers in spintronics. Femtosecond all optical switching (AOS) has revolutionized our understanding of magnetization dynamics by enabling deterministic, energy efficient magnetization reversal within sub picosecond timescales[1-8]. To date, most AOS studies have focused on spatially uniform magnetization reversal. Yet many emerging spintronic architectures also make use of non-uniform spin textures—most notably, magnetic domain walls (DWs)—as active nanoscale information carriers due to their low energy consumption, negligible current leakage, and long-term stability[9] in high-density storage devices such as magnetic random-access memory (MRAM)[10] and racetrack memory[11,12]. Extending AOS from uniform switching to the direct writing and control of DWs would therefore open new opportunities for high speed and energy efficient domain wall-based devices. However, due to the limitations of existing characterization techniques, how femtosecond laser can be used to directly write, and control magnetic domain walls remains largely unknown — posing a critical gap at the intersection of ultrafast magnetism and domain wall engineering.

Here we report the ultrafast optical writing of magnetic DWs with sub-terahertz characteristic frequencies in a prototypical ferrimagnetic GdFeCo thin film. Originally identified as the first material to exhibit all-optical switching[1], GdFeCo has since become the most extensively studied system in this field. Additionally, DWs in this system exhibit propagation speeds among the highest ever reported[13-18]—reaching several kilometers per second—enabled by the tunable angular momentum compensation temperature ($T_A$), where the sublattice angular momenta cancel. Therefore, ferrimagnetic GdFeCo offers a compelling model system for harnessing DWs on faster timescales[19]. To capture the spatiotemporal evolution of DW formation under all-optical excitation, we employ a recently developed approach that combines Lorentz ultrafast electron microscopy with transient optical grating excitation (optical TG)[20,21]. Compared to other time-resolved magnetic detection techniques—such as magneto-optical imaging techniques[22,23], X-ray Magnetic Circular Dichroism (XMCD) – Photoemission Electron Microscopy[8,24], or reciprocal-space approaches like XMCD-based resonant soft X-ray scattering[25,26]—our approach uniquely enables direct, real-space visualization of in-plane DW dynamics with sub-picosecond temporal and sub-micrometre spatial resolutions (Fig. 1a). This imaging geometry is exclusively sensitive to in-plane magnetization contrast, allowing us to directly capture the spatiotemporal evolution of ultrafast DW formation.

Fig. 1b illustrates the scope of our study. We focus on ultrafast, nonequilibrium DW nucleation dynamics driven by spatially structured laser excitation via optical TG, which induce localized quenching of magnetization. By carefully tuning the excitation fluence, we engineer periodically modulated regions of switched and unswitched magnetization within each TG period. This configuration gives rise to spatially confined in-plane magnetic transitions at the periodic switched/unswitched interfaces, where DWs nucleate and evolve. Beyond characterizing the temporal evolution of DW writing under such nonequilibrium conditions, we also reveal the spatial progression of DW formation, capturing the transition from disordered to modulated, ordered DW structures (Fig. 1c). Intriguingly, during this transition we observe strongly asymmetric, nonequilibrium Lorentz contrast signatures indicative of a transient spin texture distinct from equilibrium DWs. Multiscale micromagnetic simulations incorporating fluence-dependent local spin dynamics and alternating 180° domain environments support a nonlinear formation mechanism fundamentally different from conventional DW writing. These simulations further reproduce the transient spatially asymmetric DW features observed experimentally. Together, these findings establish a robust and tunable framework for ultrafast DW engineering and uncover a new class of far-from-equilibrium spin textures, offering a path toward dynamic control of magnetic order on ultrafast timescales.

**Spatiotemporal dynamics of ultrafast domain wall formation.** Fig. 2a presents the time-resolved evolution of Lorentz contrast following femtosecond optical TG excitation. In contrast to the raw Lorentz images shown in Fig. 1c, Fig. 2a displays Fast Fourier Transform (FFT)-filtered images[21] based on inverse FFT related to the TG spatial frequency, allowing us to isolate and highlight the magnetic contrast modulated purely by the optical TG. We observe a clear transition from spatially distorted and disordered contrast at 1 ps to the emergence of well-defined, ordered stripe-like configurations after ~7 ps. The spatial periodicity of these magnetic stripes aligns with the optical TG period, confirming that the optical TG effectively modulates the magnetic contrast. Fig. 2b shows a magnified view of the white-boxed region highlighted in Fig. 2a. To verify that this magnetic contrast originates from local ultrafast magnetization reversal, we analyzed the fluence dependence of the FFT amplitude at the TG spatial frequency at representative time delays—5 ps, corresponding to the regime dominated by nonequilibrium DW dynamics, and 87 ps, when effective field-driven ferromagnetic resonance (FMR) precession dominates. The raw data are presented in Supplementary Note 1. As shown in Fig. 2c, the FFT amplitude at 5 ps exhibits a threshold-like nonlinear dependence on fluence: no magnetic contrast is detected below ~2.4 mJ/cm$^2$, followed by a sharp increase in contrast, which subsequently diminishes in the multidomain regime due to random demagnetization[1,27,28]. In contrast, the 87 ps magnetic contrast grows monotonically with fluence, without a threshold effect. This distinction reflects the nonlinear fluence dependence of AOS[29] and confirms that the early-time magnetic contrast is specifically related to AOS, rather than other ultrafast processes such as laser-induced demagnetization, which typically lack a fluence threshold[30]. Additionally, since the sample is deliberately oriented perpendicular to the out-of-plane magnetic field (see Methods), Lorentz contrast is exclusively sensitive to in-plane magnetization gradients[20,31]. Thus, the observed early-time magnetic contrast increase can be unambiguously attributed to in-plane components in the DW formation triggered by local AOS.

We next analyzed the spatial dynamics of DW contrast formation. The Lorentz contrast evolution in Fig. 2a captures the transition from disordered in-plane magnetization to an optically modulated, spatially correlated DW grating. This increasing spatial correlation can be quantitatively assessed via the full width at half maximum (FWHM) of the FFT peak at the TG spatial frequency: a narrower peak indicates increased spatial correlation. As shown in Supplementary Note 2, the FWHM decreases by approximately 50% within ~4 ps, quantitatively confirming the rapid transition from disordered to ordered DW contrast observed in Fig. 2a. Notably, we also identify a transient strongly asymmetric magnetic contrast during the ordering process, as highlighted by the boxed region at 2 ps in Fig. 2a and shown magnified in the left panel of Fig. 2b. This strongly asymmetric contrast represents an intermediate state between the disordered and ordered DW configurations. For comparison, the right panel of Fig. 2b displays a well-established, continuous sinusoidal DW contrast at 9 ps. The transient asymmetry at 2 ps results in a second-order FFT peak at the doubled TG spatial frequency. To quantify the relationship between the two peaks, we analyzed the temporal evolution of the ratio of the second- to first-order FFT amplitudes (Supplementary Note 3). The ratio exhibits a maximum at approximately 40% around 2 ps, then rapidly decreases to about 10%, consistent with the time window in Fig. 2a during which the strongly asymmetric DW contrast magnified in the left panel in Fig. 2b is observed. This temporal evolution rules out static origins such as sample inhomogeneity, defects or systematic noise, and supports a dynamic origin related to transient in-plane spin structure transitions. To further visualize the spatiotemporal evolution of DW formation, Fig. 2d presents a space-time contour plot constructed from FFT-filtered Lorentz images in Fig. 2a. Representative horizontal line profiles extracted from this contour reveal the evolution of the DW spatial distribution. Initially, no regular contrast is observed, indicating the absence of DW formation. At 2 ps, partially developed DW contrast emerges with asymmetric "shoulder" features highlighted by the red dashed box in Fig. 2e (consistent with Figure 2b, left). These shoulder features are still present at 3.5 ps, while at later times, the line profiles evolve into uniform sinusoidal patterns attributed to Fresnel-mode defocus and confirmed by Lorentz contrast simulations (Supplementary Note 4), indicating the formation of an ordered DW configuration.

We next analyzed the temporal dynamics of DW contrast formation. As shown in Fig. 2f, the FFT amplitude at the DW grating frequency rises rapidly and is well described by an exponential function with a time constant of 2.2 ± 0.6 ps, saturating within ~10 ps. This corresponds to a sub-terahertz DW writing speed. Superimposed on this rise is a coherent oscillation with a frequency of 0.25 THz (Fig. 2g). This sub-THz mode corresponds to an exchange-mode spin resonance[32], where the magnetization precesses about the inter-sublattice exchange field. This exchange field, arising from the quantum mechanical coupling between rare-earth and transition-metal sublattices, has an effective field strength on the order of ~10 T, enabling precession frequencies far beyond conventional FMR. Atomistic spin dynamics simulations incorporating sublattice exchange interactions reproduce this frequency range (Supplementary Note 5). As illustrated in the macrospin model in Fig. 2h, the origin of the DW contrast oscillation is attributed to the transient canting angle induced by different demagnetization and reversal rates of the sublattices, producing DW oscillations driven by the inter-sublattice exchange field. This sublattice exchange interaction has previously been implicated in AOS phenomena such as transient ferromagnetic-like states[2], determination of switching timescales[33], and dominance in dissipative relaxation process[34,35]. Here, we show that it also plays a direct role in ultrafast DW formation.

Together, the transient strongly asymmetric 'shoulder' spatial features and the sub-terahertz oscillatory dynamics observed during ultrafast DW formation reveal a previously unrecognized metastable DW state. In particular, the spatial evolution of DW contrast highlights a nonequilibrium formation pathway, indicating that DW writing under all-optical excitation fundamentally lies beyond the scope of spatially averaged, quasi-equilibrium models.

**Fluence-dependent manipulation of nonequilibrium domain wall states and textures**. To explore the control of DW dynamics by all-optical excitation, we performed a spatially resolved analysis of fluence-dependent DW behavior. Interestingly, beyond observing the two states where metasable DWs were written and not written, we also discovered a third intermediate state where metasable DWs are written transiently. By scanning the sample (~30 × 30 μm$^2$, Fig. 3a) under a Gaussian laser profile, we identified three distinct DW states: no DW formation (white regions), transient DWs (orange regions), and metastable DWs (blue regions). The transient DW state exhibits a rapid rise and decay in Lorentz contrast within ~10 ps, indicating rapid recovery of DW contrast (orange curve in Fig. 3b). As shown in the measured spatial fluence distribution in Supplementary Note 6, regions without the formation of DWs emerge in areas with lower fluence, while metastable DWs appear in regions with relatively higher fluence. Transient DW states are observed in the intermediate zones transitioning from low to high fluence, confirming the fluence-dependent nature of this behavior and highlighting the transitional character of the transient DW dynamics. The appearance of such transitional transient DW dynamics is noteworthy and calls for deeper physical insight. As previously noted, the current experimental setup does not allow for the observation of demagnetization contrast. Therefore, the emergence of transient DW contrast, as indicated by the orange curve in Fig. 3b, suggests the formation of a transient in-plane spin structure, prior to the formation of metastable DW in higher fluences.

Based on the fluence-dependent dynamics observed in both Fig. 2c and Fig. 3a–b, we identified four distinct DW regimes as fluence increases (Fig. 3c): (I) no DW formation due to insufficient fluence for triggering local AOS; (II) transient DW recover spontaneously and rapidly on a timescale of ~10 ps; (III) metastable DWs persisting beyond tens of picoseconds (Fig. 2f); and (IV) multidomain DW textures associated with high-fluence excitation.

We further increased the background excitation (or the mean fluence of the optical TG), driving the sample into a multidomain state. Interestingly, we observed the formation of anti-phase DW textures. Upon TG excitation across an initial static DW formed in a multidomain state (Fig. 3d), Lorentz imaging reveals a periodic reversal of magnetization across the static DW (Fig. 3e-g), forming a spatially alternating contrast pattern (an antiphase domain boundary). Time-resolved measurements show a ~500 ps jump in recovery time across the initial DW position (Fig. 3h-j). This asymmetry originates from the anisotropy field $\mathbf{H}_{ani}$, which adds to or opposes the external field $\mathbf{H}_{ext}$ on either side of the permanent domain wall, thereby modifying the effective field $\mathbf{H}_{eff}$ that ultimately governs the system's recovery time[36]. Detailed determination, and mechanisms of recovery dynamics are provided in Supplementary Note 7. These findings demonstrate that optical excitation alone enables tunable DW textures and lifetimes.

**Multiscale simulation of nonlinear domain wall formation via localized spin textures.** Fig. 3i shows the metastable DW recovery dynamics dominated by the effective field, lasting for hundreds of picoseconds, which is consistent with the recovery mechanism reported in pump–probe AOS experiments in the literature[36]. This stands in stark contrast to the unconventional picosecond-scale spontaneous recovery of the transient DW state, as shown by the orange curve in Fig. 3b. To further understand the microscopic origin of the short-lived in-plane spin structure appearing during the fluence transition in Fig. 3b, as well as the transient asymmetric DW contrast with a "shoulder" feature observed in Fig. 2—occurring as an intermediate state between disordered and ordered DW contrast, we employed a multiscale micromagnetic simulation approach[26] that combines atomic spin dynamics (ASD) with micromagnetic simulations to investigate the spatiotemporal dynamics of local AOS-induced domain walls under optical TG excitation. Fig. 4a depicts the sinusoidal optical TG fluence distribution (period: 1 µm), mimicking experimental excitation. Fluence values sampled along this profile were input into ASD simulations of a $100 \times 100 \times 5$ nm$^3$ GdFeCo region. Fig. 4b presents the magnetization snapshots at 3 ps delay for representative fluences. At low fluence, only ultrafast demagnetization occurs (white regions), whereas higher fluences trigger AOS (black regions), resulting in magnetization reversal. The spin structures are highly non-uniform across fluences (colored regions represent the in-plane spin polarization, depicted by the color wheel in the upper right corner of the leftmost panel in Fig. 3b), revealing nanoscopic localized spin textures—termed dynamical magnon bound states, or magnon drops[26,37-40]. The detailed ASD results at all fluence levels are presented in Supplementary Note 8. Interestingly, the fluence-dependent ASD data reveal that at the threshold fluence for local AOS, the localized spin textures favoring in-plane magnetic configurations exhibit larger sizes (middle panel of Fig. 4b) and longer lifetimes (Supplementary Note 8). Since the domain wall nucleation sites mediated by the optical TG lie within this threshold fluence range, the ultrafast nucleation of domain walls originates from the transformation of these discrete, unstable localized in-plane spin textures.

In the experimental setup, the actual environment for DW nucleation is influenced not only by the threshold fluence but also by exchange interactions arising from the alternating spin-up and spin-down 180° domain configurations caused by local magnetic reversal. To account for both effects, we employed micromagnetic simulations initialized with ASD-generated magnetization profiles tiled across the TG fluence landscape. We plotted the results obtained at all time steps into a space-time contour, as shown in Fig. 4c. The upper panel of Fig. 4c presents a complete transverse slice of the space-time contour at 4 ps. From the space-time contour of a single TG period derived from the simulation, it is evident that at the initial moment, the color variations representing the in-plane magnetic polarization are highly diffuse and irregular. This indicates a spatially disordered spin configuration in the DW nucleation regions, originating from the formation of localized spin textures induced by threshold fluence, as confirmed by ASD simulations. These localized and non-uniform in-plane spin structures reflect a strongly non-equilibrium spin state immediately following optical excitation. As time progresses, driven primarily by exchange interactions, the system gradually evolves into a continuous and well-defined DW structure. The DW profile becomes progressively sharper, indicating a transition toward a spatially ordered spin configuration. This process exemplifies a clear case of non-equilibrium spin texture evolution and characterizes the relaxation dynamics of the DW toward a lower-energy and more ordered magnetic state. During this transition from disorder to order, the system passes through an intermediate state, in which localized, disordered spin textures coexist with partially developed DW segments, as illustrated in the upper panel of Fig. 4c. This representative simulation snapshot shows that the DW region is dominated by disordered localized spin textures, while the edges display partially formed and slightly curved DW segments. These features highlight the coexistence of non-equilibrium spin fluctuations and the emergence of magnetic order, supporting the concept of a hybrid spin configuration during the relaxation process. This hybrid configuration reflects a transient competition between local non-equilibrium fluctuations and the global drive toward an energetically favorable DW pattern.

Furthermore, we simulated the Lorentz contrast corresponding to this intermediate-state spin configuration within a single TG period, as shown in the upper-left panel of Fig. 4d. The simulated Lorentz image reveals a pronounced shoulder feature, which is consistent with the transient shoulder contrast observed in experiments (Fig. 2b, 2e, and the lower-left panel of Fig. 4d). In contrast, both the simulated and experimentally observed fully developed, ordered DW structures exhibit a smooth, continuous sinusoidal distribution without shoulder features (Fig. 2b, 2e, and right panel of Fig. 4d).

Additionally, to quantitatively explain the temporal features of DW growth, we calculated the time evolution of the spatial profile of total in-plane magnetization component $M_y$ along the *x*-axis from the micromagnetic simualtions (Fig. 4e) since the Lorentz contrast in our experiment is sensitive to the gradient of $M_y$ along the *x* direction. Initially, $M_y$ exhibits discrete fluctuations reflective of isolated localized spin textures. Over time, these features evolve into a continuous distribution, indicative of metastable DW formation. FFT analysis of this $M_y$ profile yields a temporal growth curve of the FFT intensity at the dominant DW spatial frequency (Fig. 4f). The FFT intensity exhibits exponential growth, from which we extract a characteristic DW formation time constant of 3.6±0.5 ps, closely matching the experimental value of 2.2±0.6 ps (Fig. 2). The small discrepancy likely stems from the discrete nature of the simulated fluence distribution and the exclusion of exchange flow spin currents (EFSCs), which are known to accelerate spin dynamics[26,41] and could improve agreement with experiment.

**Outlook**

Our results establish ultrafast all-optical DW writing as a fundamentally efficient approach to sub-terahertz spintronic control. We observe DW writing dynamics with an exponential rise time constant of ~2 ps —nearly two or even three orders of magnitude faster than conventional techniques such as field-driven[42], spin-transfer torque (STT)[43], spin–orbit torque (SOT)[44], or strain-mediated methods[45]. These all-optically written domain walls arise from an ultrafast, thermally mediated quenching–remagnetization process, representing highly non-equilibrium magnetisation dynamics and reveals a previously unidentified metastable domain wall nucleation pathway that proceeds via a transition state between disordered and ordered phases, fundamentally distinct from known mechanisms. Multiscale micromagnetic simulations support this picture by showing that the ultrafast transition is governed by the evolution from disordered, discrete localized spin textures to an ordered and continuous domain wall, providing a plausible explanation for the observed spatiotemporal dynamics. Therefore, our findings not only demonstrate a new paradigm for high-speed DW engineering but also uncover a distinct DW formation mechanism far from equilibrium. It is reasonable to expect that these intermediate transitional DW states may serve as a precursor platform for realizing emergent classes of DW–embedded topological spin textures. Recent studies have experimentally identified DW–embedded topological defects—such as DW bimerons[46] and DW skyrmions—not only as promising candidates for topologically protected information carriers, but also as fundamentally new spin textures with unique internal topology and spatial confinement. We speculate that under appropriate inversion symmetry breaking or in the presence of interfacial Dzyaloshinskii–Moriya interaction (DMI), the nonequilibrium, localized spin texture-mediated DWs observed in our study may evolve into various topological configurations in a controlled manner, potentially enabling new routes for the creation and manipulation of confined spin textures under optical excitation. Looking ahead, our findings also open up several avenues for future exploration—including nanoscale imaging of metastable DW formation, investigation of potential extreme DW velocities under optical excitation, interactions between exchange-mode magnons and ultrafast formed DWs, and superdiffusive spin current contribution to ultrafast DW formation. These directions may extend the frontier of ultrafast spintronics by harnessing nonequilibrium magnetic textures on sub-picosecond and nanometre scales.

## Methods

### Sample geometry and preparation

Thin films of $Gd_{0.24}(FeCo)_{0.76}$ were fabricated on 5 nm tantalum seed layer coated electron-transparent $Si_3N_4$ membranes (Ted Pella, USA) with a thickness of 50 nm. The window dimensions of these supporting grids measured 150 μm × 150 μm. $Gd_{0.24}(FeCo)_{0.76}$ films were sputtered via magnetron sputtering system using argon as a process gas, using high-purity (≥99.95%) Gd and $Fe_{50}Co_{50}$ targets. The chamber base pressure prior to deposition was maintained below $2 \times 10^{-8}$ Torr. To inhibit surface oxidation, a protective 5 nm tantalum layer was subsequently deposited on top of the magnetic layer.

### Lorentz mode ultrafast electron microscopy (LUEM)

LUEM based on a modified JEOL JEM-2100 transmission electron microscope was employed to investigate the real-space dynamics of DW formation induced by local AOS in $Gd_{0.24}(FeCo)_{0.76}$. A fiber-based femtosecond laser system (Tangerine, Amplitude Systèmes) delivering 1030 nm pump pulses of ~300 fs duration at a 12 kHz repetition rate served as the optical pump source. The optical TG was formed directly on the sample by interfering two linearly polarized pump beams: one incident beam and a secondary beam reflected from an independent slanted face angled at 35° relative to the membrane normal, creating a sinusoidal spatial modulation of laser fluence on the film surface[20]. This optical interference pattern gives rise to a spatially periodic distribution of local AOS within each grating period. As a consequence of this localized switching, magnetic DWs form at the boundaries between the switched and unswitched regions.

The electron pulses used to probe the ultrafast magnetic response were generated by photoemission from a $LaB_6$ cathode using the fourth harmonic (258 nm) of the same laser source. These electron pulses had a duration of ~1.2 ps. The objective lens was kept off during magnetic imaging using the free lens control mode, with external magnetic fields introduced by carefully applying a calibrated current to the lens coils. This setup allowed precise control of the out-of-plane magnetic field ($H_{ext}$) required to reset the spin system between pump pulses in the stroboscopic pump-probe acquisition. For the experiments discussed here, the sample was measured without tilt, ensuring that the applied magnetic field was strictly perpendicular to the film plane. Under this alignment, the observed Lorentz contrast originates predominantly from in-plane magnetization gradients, allowing the visualization of an increase in in-plane magnetic contrast purely driven by DW nucleation dynamics during the first tens of picoseconds, before the onset of FMR-mode precession. Ultrafast Lorentz images were acquired using a CheeTah T3 detector (Amsterdam Scientific Instruments).

**Atomistic spin dynamics (ASD) simulations**

Ultrafast magnetic dynamics were modeled using atomistic spin simulations performed on $Gd_{0.24}(FeCo)_{0.76}$ alloy, capturing key ultrafast phenomena such as ultrafast demagnetization and AOS dynamics. Time-dependent spin dynamics at each atomistic site is governed by the stochastic Landau–Lifshitz–Gilbert equation with Langevin dynamics. The Hamiltonian includes nearest-neighbor exchange interactions and uniaxial anisotropy. Both ferromagnetic (e.g., FeCo–FeCo, Gd–Gd) and antiferromagnetic (e.g., FeCo–Gd) couplings were included to reflect the alloy's sublattice structure. Laser-induced heating was introduced via Langevin dynamics, with temperature evolution derived from a two-temperature model (TTM)[47].

VAMPIRE software package[48] was used to perform the atomistic spin simulations, where the stochastic equations were integrated via the Heun numerical integration method. The local atomic moments were set to $1.92\mu_B$ for FeCo and $7.63\mu_B$ for Gd, with $\mu_B$ denoting the Bohr magneton. Parameters denoting magnetic exchange couplings were $J_{FeCo-FeCo}=2.835\times10^{-21}$J, $J_{Gd-Gd}=1.26\times10^{-21}$J, and $J_{FeCo-Gd}=-1.09\times10^{-21}$J. A uniaxial anisotropy energy of $8.07\times10^{-24}$ J per atom was applied. Simulations were carried out with lateral dimensions of 100×100 nm², with a thickness of 5 nm.

**Multiscale micromagnetic simulations**

To elucidate the spatiotemporal evolution of AOS induced DW formation dynamics, we employed a multiscale simulation framework that integrates atomistic spin dynamics simulations with micromagnetic simulations. ASD simulations under various fluences were individually performed over a $100 \times 100 \times 5$ nm³ region to resolve the localized ultrafast magnetization dynamics governed by the spatial fluence distribution imposed by optical TG. The set of fluences sampled for the ASD simulations spans both below and above the critical AOS threshold fluence, ensuring that the transition across the threshold is fully captured. The resulting magnetization maps at a delay of 3 ps, served as the initial condition for the subsequent micromagnetic simulations. We chose the 3 ps magnetization state from ASD result as the initial condition for the micromagnetic simulations because, after the critical 3 ps, the system has evolved beyond the sublattice antiferromagnetic-coupling–dominated dynamics (see Supplementary Note 9) into a stabilized effective ferromagnetic state that can be consistently described by a micromagnetic model[26].

The ASD results at 3 ps, each associated with a specific fluence, were combined according to the fluence profile of the optical TG, using uniformly sampled fluence values across a single TG period. The combined dataset is then used as input for micromagnetic simulation. The total simulated region spans 1200 nm along the *x*-axis and 100 nm along the *y*-axis, corresponding to one TG period. The long-term relaxation of the entire system within a single TG period, where local AOS occurred, was then modeled using the the graphic processing unit (GPU) package MuMax3[49], which solves the Landau–Lifshitz equation for a continuous magnetic system. We performed the micromagnetic simulation using the cell size of $1 \times 1 \times 1$ nm$^3$ optimized to balance resolution and computational efficiency while respecting the exchange length. The grid size of $1200 \times 1200 \times 5$ nm$^3$ was used. Material-specific parameters were set as follows: saturation magnetization $M_s$=2e$^5$ A m$^{-1}$, uniaxial anisotropy constant $k_u$=4e$^4$ J m$^{-2}$, exchange stiffness $A$=1e$^{-11}$ J m$^{-1}$, and Gilbert damping $\alpha$=0.1. Micromagnetic simulations were run on NVIDIA GPUs using native CUDA implementation under periodic boundary conditions. The effective field terms included exchange, anisotropy, and external magnetic field contributions. Temporal integration of the LLG equation was performed using a default Mumax3 solver. This multiscale framework enabled us to effectively connect the ultrafast, localized spin texture dynamics resolved by ASD simulations with the subsequent metastable continuous DW formation governed by exchange-driven evolution resolved by micromagnetic simulations. Crucially, it allowed us to uncover the transformation pathway from transient, unstable localized spin textures into long-lived and metastable DW configurations.

**Lorentz contrast simulation**

To simulate the Lorentz contrast, we employed the PyLorentz open-source software package[50] to perform electron phase shift calculations incorporating both magnetic vector and electrostatic contributions, given by the Aharonov-Bohm relation. Starting from micromagnetic outputs, the spatially resolved magnetization was discretized into finite elements, and their individual potential contributions were combined linearly to obtain the total phase shift distribution. The resulting electron exit wavefunction was then convolved with a microscope-specific transfer function to obtain the final image wave function, which accounts for instrumental parameters such as defocus, astigmatism, spherical aberration, and beam coherence. This procedure enables direct generation of simulated Lorentz images from the multiscale micromagnetic simulations results.

**Acknowledgments**

**Funding:** This research was funded by the Knut and Alice Wallenberg Foundation (2012.0321 and 2018.0104), the Swedish Research Council (VR), and through the ARTEMI national infrastructure (VR 2021-00171 and Strategic Research Council (SSF) RIF21-0026). JL W acknowledge National Natural Science Foundation of China (12241403, 12304136), Guangdong Basic and Applied Basic Research Foundation (Grant Nos. 2022A 1515110863 and 2023A1515010837), Guangdong Science and Technology Project (grant no. 2023QN10X275) and Project for Maiden Voyage of Guangzhou Basic and Applied Basic Research Scheme (Grant No. 2024A04J4186) for the computing resources. YF acknowledge Yichen Wu for the meaningful discussion during writing the paper. **Competing interests:** The authors declare that they have no competing interests. **Data and materials availability:** All data needed to evaluate the conclusions in the paper are present in the paper and/or the Supplementary Materials.

**Author contributions**

J.W. supervised the project. G.C. proposed the initial idea. Y.F., G.C., and J.W. designed the experiments. Y.F. and G.C. performed the experiments.  J.S. and J. Å. fabricated the sample. JL. W. performed the atomistic spin model simulation. Y.F. performed the micromagnetic simulations and Lorentz contrast simulation. Y.F. and J.W. wrote the paper. All authors participated in reviewing the paper and offering feedback.

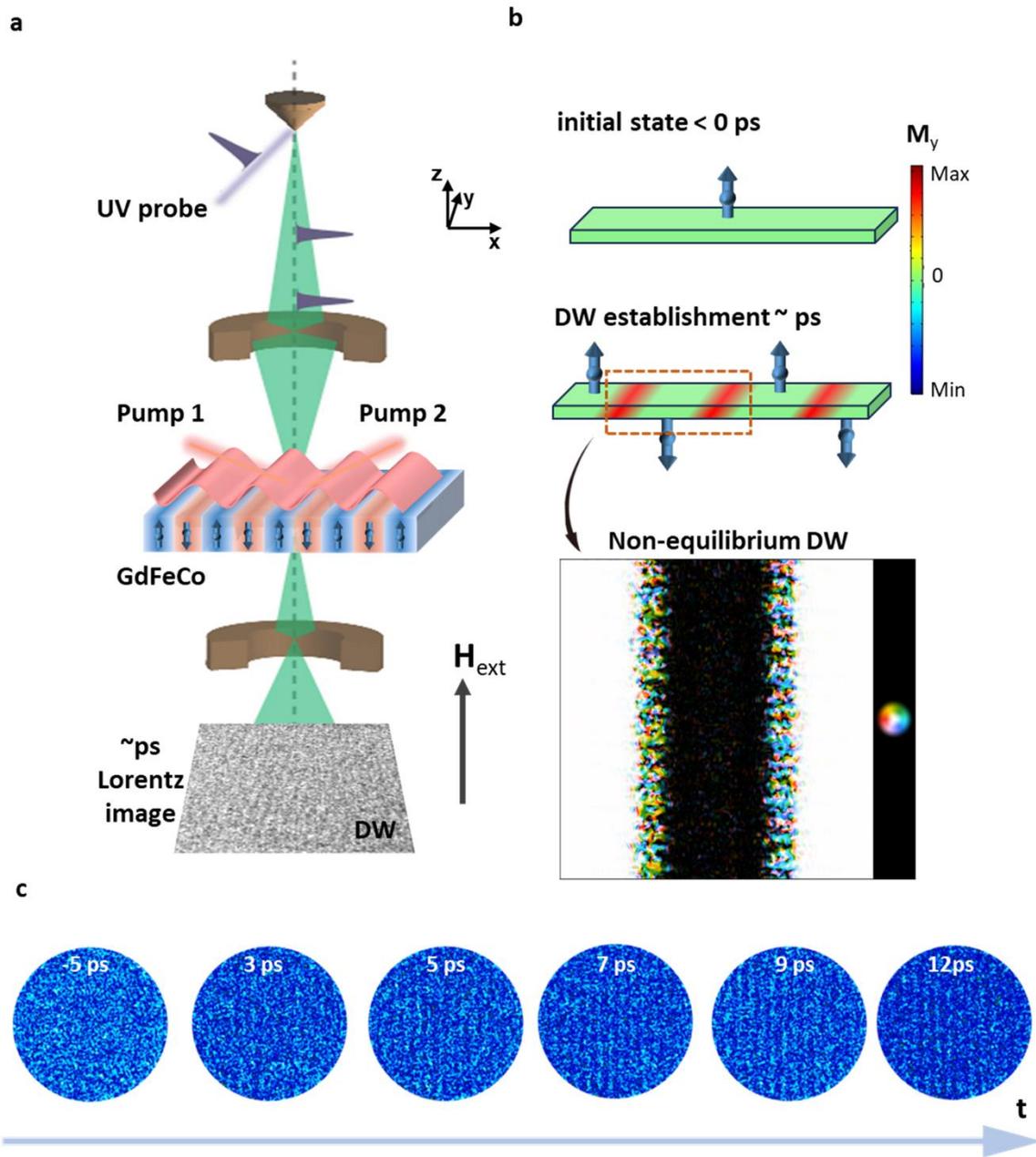

Figure 1: **Schematic of Experimental Setup and Evidence for AOS-Induced DW Contrast.** (a) Schematic of the Lorentz ultrafast electron microscope (LUEM) combined with transient optical grating (TG) excitation. A UV probe beam illuminates the cathode, generating ultrafast electron pulses, while two interfering IR pump beams create an optical TG with sinusoidal spatial modulation of laser fluence. When this optical TG is applied to the GdFeCo film, regions of constructive interference undergo localized AOS within a few picoseconds, reversing the magnetization (red areas), while regions of destructive interference retain the initial state (blue areas), resulting in a periodic alternating 180° spin up and down domains pattern. The LUEM contrast is sensitive to in-plane magnetization gradients. The bottom panel shows a raw Lorentz image at 5 ps, revealing a characteristic metastable DW grating. An out-of-plane external magnetic field is applied to reset the system between pump-probe cycles. During the experiment, the sample was kept at a tilt angle of 0°, maintaining a strictly perpendicular orientation to the external magnetic field. (b) The upper two panels (illustrations) show DW nucleation forms within locally switched/unswitched in-plane transitional regions, where green indicates a Lorentz contrast of zero. The lower panel is a schematic representation based on simulation data, illustrating the typical non-equilibrium DW structure formed after laser quenching. (c) Temporal evolution of representative raw LUEM results. From left to right, the corresponding time delays are –5 ps, 3 ps, 5 ps, 7 ps, 9 ps, and 12 ps. Full results are available in Supplementary Video 1.

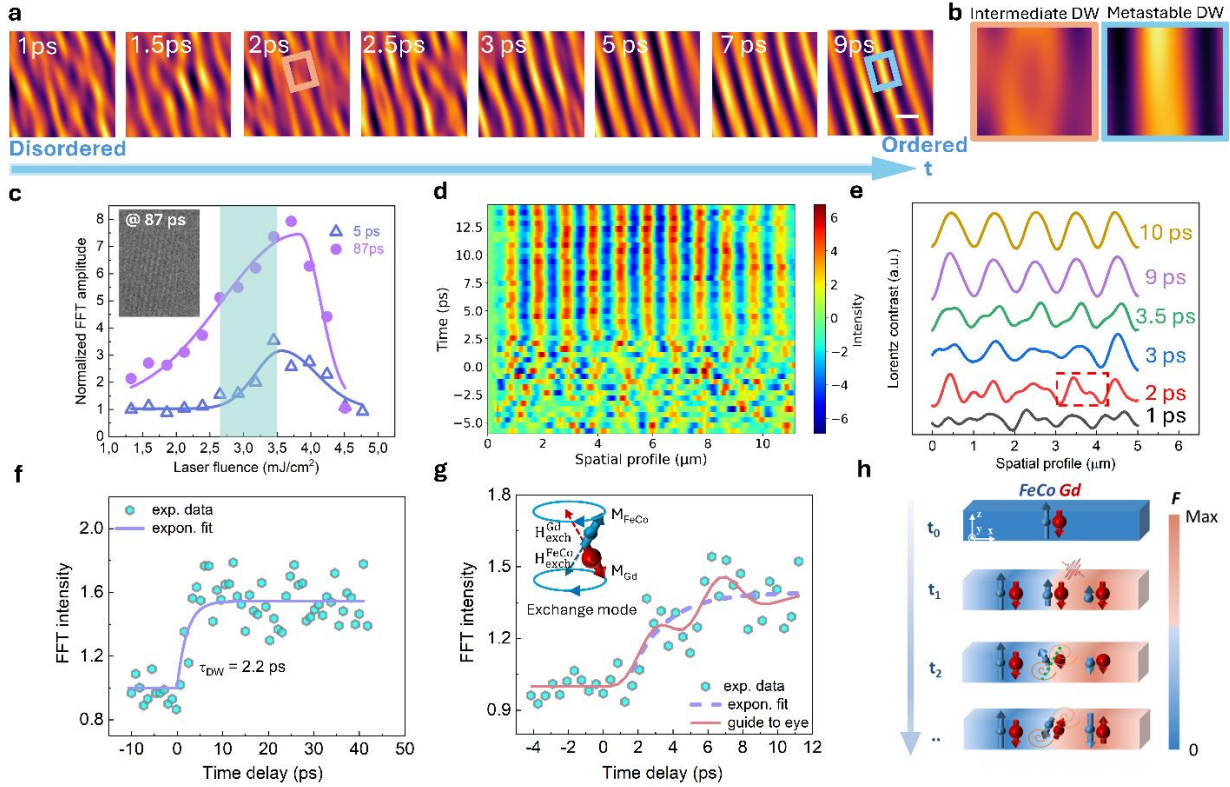

Figure 2: **Spatiotemporal observation of ultrafast DW formation dynamics.** (a) FFT filtered Lorentz images at progressive time delays, revealing the evolution of the DW pattern in real space from a disordered to an ordered state. FFT-filtered Lorentz images were reconstructed by applying masks at the TG-related signal peaks in the Fourier domain, followed by inverse FFT. The mask radius was determined by the full width at half maximum (FWHM) of the peaks. The scale bar represents 1 μm. Full results are available in Supplementary Video 2. (b) An enlarged view of the representative DW contrast morphology highlighted by the white boxes in (a). The left panel displays a transient DW contrast with pronounced asymmetry at 2 ps, while the right panel shows a well-established, continuous, and ordered metastable DW contrast at 9 ps. (c) Fluence-dependent magnetic contrast at different typical processes. Magnetic contrast dominated by DW formation process (at 5 ps) and by ferromagnetic resonance precession process (at 87 ps) shows different fluence responses. At 5 ps, magnetic contrast exhibits a clear fluence threshold, confirming the presence of AOS-triggered DW contrast. In contrast, at 87 ps, the magnetic contrast increases smoothly without a threshold. (d) Space-time contour of the evolution of patterned DW contrast, obtained by extracting transverse line profiles perpendicular to the grating direction from the FFT-filtered images of (a). (e) Line profiles extracted at representative time delays from the space-time contour shown in (d) reveal the evolution of the DW contrast—from an initially disordered state at 1 ps, to the emergence of pronounced shoulder features at 2–3.5 ps (highlighted by the red box), and finally to the formation of a well-established, ordered sinusoidal DW pattern. (f) Normalized FFT intensity as a function of time delay within 40 ps. The purple line represents a single-exponential fit, yielding a time constant of 2.2 ± 0.6 ps. The FFT intensity corresponds to the intensity at the optical

TG spatial frequency, positively correlated to the DW contrast. (g) Normalized FFT intensity as a function of time delay within 8 ps. The purple curve shows the exponential fit, while the red line serves as a guide to the eye, highlighting oscillations superimposed on the exponential growth. The inset shows schematic illustration of exchange mode precession around the sublattice antiferromagnetic exchange field. (h) Schematic of the macrospin model, in which each sublattice is simplified to a single spin, shown at a single TG period after TG excitation. The color bar represents the fluence distribution, with red indicating regions of constructive interference and blue indicating destructive interference regions. Due to the distinct demagnetization and reversal rates of different sublattices, a non-collinear canting angle arises between sublattice spins, accompanied by exchange mode oscillations associated with the formation of metastable magnetic DWs.

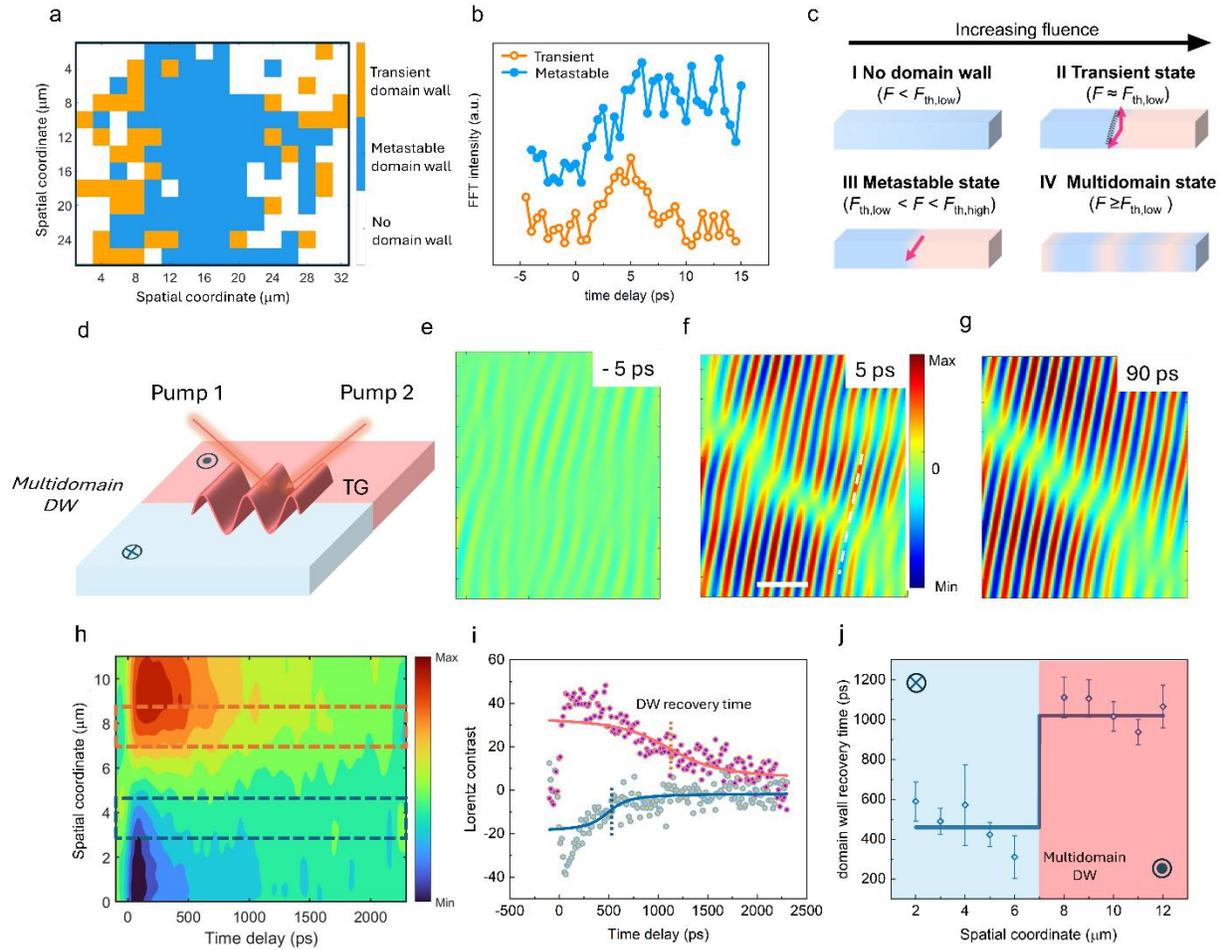

Figure 3: **Fluence-dependent manipulation of non-equilibrium DW states and textures.** (a) Spatial map of local DW dynamics under Gaussian laser excitation: white, orange, and blue regions correspond to no DW formation, short-lived DWs (spontaneously recovers within ~10 ps), and metastable DWs, respectively. The identification of different DW dynamics is determined based on the temporal evolution of DW contrast in localized areas, as shown in (b). (b) The characteristic curves of FFT intensity as dependent on time delay at representative local regions in (a). The blue and orange lines represent the typical DW dynamics in the corresponding blue and orange regions of (a), respectively. (c) Schematic illustration of four DW states within one grating period (Types I–IV) as fluence increases: Type I, where DWs cannot be written by insufficient fluences to trigger local AOS; type II, indicating a short-lived transitional DW state with spontaneous recovering within ~10 ps at transitional fluences; type III, representing the metastable DW writing; and type IV, multidomain DW textures appear associated with enhanced mean fluence of optical TG. (d) Schematic of optical TG excitation across an initial static 180° DW formed within a multidomain state induced by high-fluence excitation. (e–g) Lorentz images collected at -5 ps, 5 ps, and 90 ps, reveal the emergence of anti-phase DW textures. All Lorentz images have been FFT-filtered to highlight the transient magnetic gratings. The color bar represents the magnitude of the Lorentz

contrast, with blue and red indicating opposite signs and green representing zero contrast. (h) Space-time contour along the white dashed line in (f). The color bar indicates the magnitude of the Lorentz contrast. (i) Typical Lorentz contrast as a function of time, from the regions marked by the white boxes in (h). The dashed vertical lines show the fitted DW recovery times. The opposite signs of the Lorentz contrast reflect opposite directions of magnetization. (j) The DW recovery time as a function of spatial coordinate, derived from the time-dependent Lorentz contrast curves at various spatial regions in (h). The boundary between the light blue and red areas in the background indicates the position of the initial 180° DW, consistent with the schematic in (d). Error bars represent the standard errors obtained from the fitting procedure.

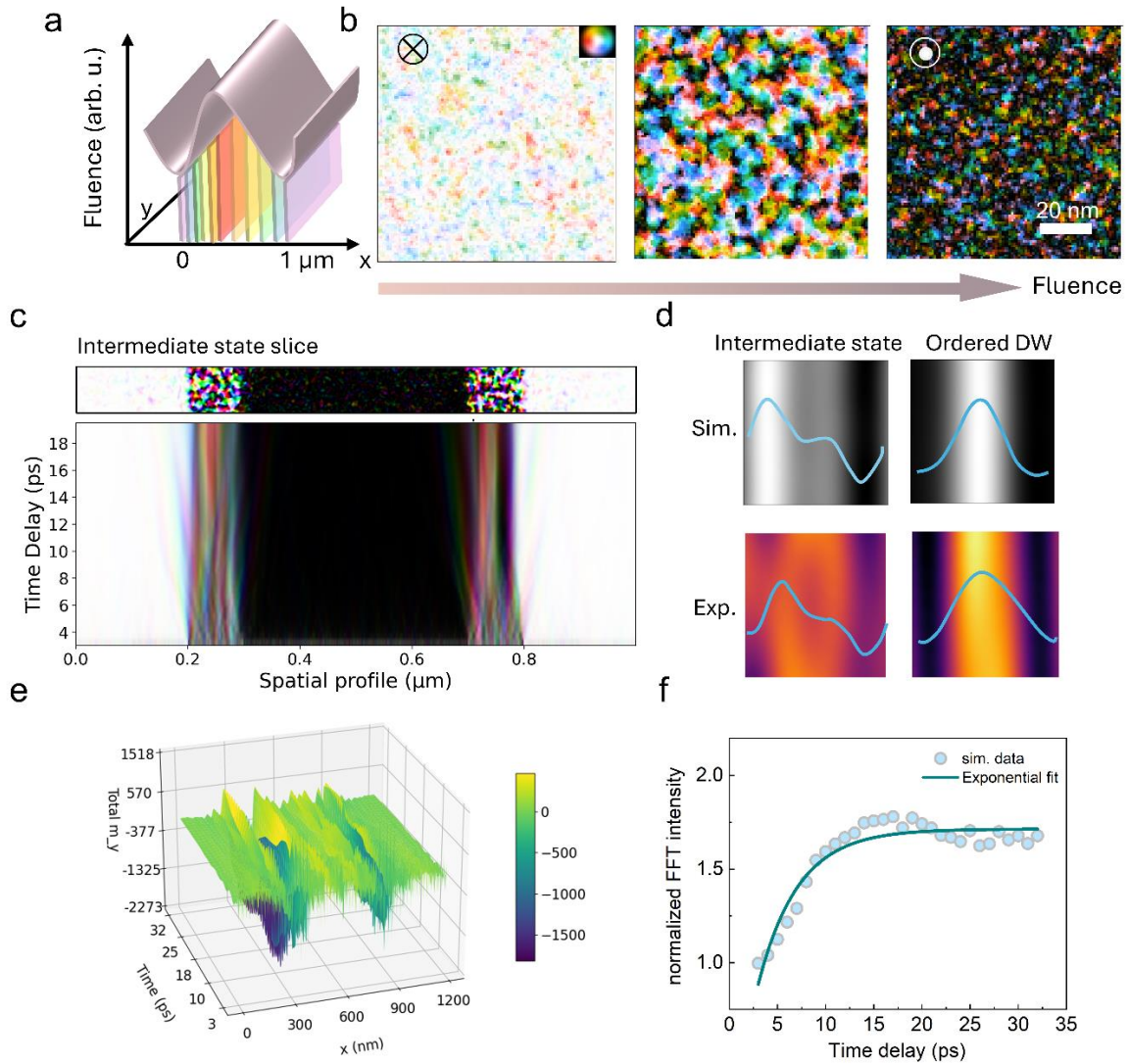

Figure 4: **Multiscale simulation of nonlinear domain wall formation via localized spin textures.** (a) Sinusoidal optical TG fluence profile across a single period, used as the input excitation landscape for simulations. Ten fluence levels are uniformly sampled for atomistic modeling. (b) Snapshots from atomistic spin dynamics (ASD) simulations at 3 ps show spin configurations under increasing fluence. Low fluence leads to ultrafast demagnetization (white), while higher fluence induces local all-optical switching (black), accompanied by the emergence of unstable, nanoscopic spin textures (magnon drops). The in-plane spin polarization is color-coded by direction (see wheel in the upper right corner), revealing nonuniform localized structures that serve as nucleation precursors. Full ASD results are available in Supplementary Video 3. (c) Multiscale micromagnetic simulations initialized with ASD-generated profiles tiled across the TG fluence landscape. The subsequent micromagnetic simulations incorporated the effects of exchange and anisotropy fields, external magnetic field and Gilbert damping, capturing the system's temporal evolution. The upper panel shows a snapshot of the full multiscale micromagnetic simulation result at representative 4 ps,

revealing an intermediate state that features both disordered localized spin textures and partially ordered DW edges. Full snapshots are available in Supplementary Video 4. The results at all time delays are summarized in the space-time contour within a single TG period shown in the lower panel. The color represents the in-plane spin polarization, illustrating the evolution of DW structure from a disordered state with a diffuse color distribution to an ordered DW state characterized by sharp color distribution at the DW regions. (d) In the representative intermediate DW state (2 ps, left column) and the ordered DW state (20 ps, right column), Lorentz images of a single TG period simulated from the multiscale micromagnetic results in (c) (upper panel) are compared with the corresponding experimental Lorentz images (lower panel). The simulation successfully reproduces the strongly asymmetric "shoulder" structure observed transiently in Fig. 2. (e) Spatiotemporal evolution of the integrated in-plane magnetization component $M_y(x,t)$ showing the transition from discrete to continuous DW structures. (f) FFT analysis was performed on the total $M_y(x,t)$ profile in (e), extracting the evolution of FFT intensity at spatial frequencies corresponding to DW positions. The resulting contrast shows exponential growth, and fitting the time-dependent curve (blue data) with a single-exponential function yields a time constant of $3.6 \pm 0.5$ ps.